\documentstyle[amssymb,prl,aps,epsfig]{revtex}

\begin{document}
\title{The mystery of superconductivity in the cuprates evinced by
London penetration depths measurements}
\author{T. Schneider}
\address{Physik-Institut der Universit\"{a}t Z\"{u}rich, Winterthurerstrasse 190,\\
CH-8057 Z\"{u}rich, Switzerland}
\maketitle
\bigskip

\begin{abstract}
The London penetration depth plays a key role in determining and
uncovering many properties of a superconductor, including
homogeneity, anisotropy, isotope effects, importance of quantum
and thermal fluctuations, and facets of the nature of
superconductivity in a particular material. Guided by the generic
phase diagram in the temperature-dopant concentration plane we
examine experimental data on the temperature, isotope
substitution, inhomogeneity and magnetic field dependence of the
penetration depths to uncover some facets of the mystery of
superconductivity in the cuprates.
\bigskip

\end{abstract}


\bigskip
To appear in Proceedings of the International School of Physics,
Enrico Fermi Course CLV, The Physics of Complex Systems.

\bigskip

\bigskip

\begin{figure}[tbp]
\centering
\includegraphics[totalheight=6cm]{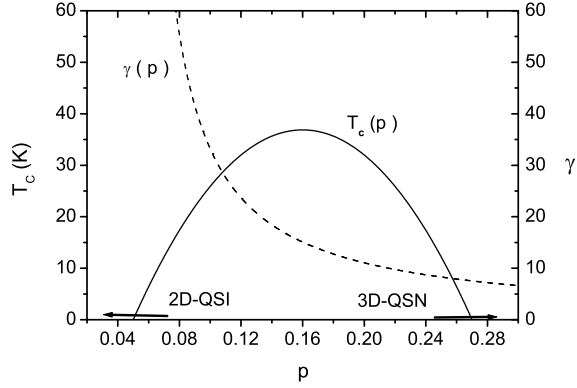}
\caption{Schematic phase diagram of cuprate superconductors. Variation of $%
T_{c}$ and $\protect\gamma \left( T=0\right) =\protect\lambda _{c}\left(
T=0\right) /\protect\lambda _{ab}\left( T=0\right) $ with hole concentration
$p$. Quantum phase transitions occur at the endpoints $p=p_{u}\simeq 0.05$
and $p=p_{o}\simeq 0.27$ of the critical line $T_{c}\left( p\right) $. At $%
p_{u}$ a two dimensional quantum superconductor to insulator (2D-QSI)- and
at $p_{o}\simeq 0.27$ 3D quantum superconductor to normal state (3D-QSN)-
transition occurs.}
\label{fig1}
\end{figure}

The London penetration depth is a fundamental quantity of a
superconductor. It plays a key role in determining and uncovering
many properties of a superconductor, including homogeneity,
anisotropy, isotope effects, importance of quantum and thermal
fluctuations, and facets of the nature of superconductivity in a
particular material. In recent years experimental data on the
temperature, dopant concentration, magnetic field and oxygen
isotope mass dependence of the penetration depth became available
for a variety of cuprate superconductors. Here we analyze and
discuss the experimental data, guided by the generic phase diagram
of the cuprates,
depicted in Fig.\ref{fig1}. After passing the so called underdoped limit $%
\left( p_{u}\approx 0.05\right) $,\ where $p$ is the hole concentration, $%
T_{c}$ reaches its maximum value $T_{c}^{m}$ at $p_{m}\approx
0.16$. With further increase of $p$, $T_{c}$ decreases and finally
vanishes in the overdoped limit $p_{o}\approx
0.27$\cite{tallon,presland}.\ There is the line $T_{c}\left(
p\right) $ of finite temperature phase transitions, separating the
superconducting and non-superconducting states, with critical
endpoints at $p_{u}$ and $p_{o}$ . Here $T_{c}$ vanishes and the
cuprates undergo at zero temperature doping ($p$) tuned quantum
phase transitions. As their nature is concerned, resistivity
measurements reveal a quantum
superconductor to insulator (QSI) transition in the underdoped limit ($p_{u}$%
) and in the overdoped limit ($p_{u}$) a quantum superconductor to
normal state (QSN) transition. Another essential experimental fact
is the doping dependence of the anisotropy. In tetragonal cuprates
it is defined as the ratio $\gamma =\lambda _{c}/\lambda _{ab}$ of
the London penetration depths
due to supercurrents flowing perpendicular ($\lambda _{c}$ ) and parallel ($%
\lambda _{ab}$ ) to the ab-planes. Approaching the QSN transition
$\gamma $ remains finite, while at the QSI transition it tends to
infinity\cite {tseuro,tsphysB}. When $\gamma $ remains finite the
system exhibits anisotropic but genuine three dimensional (3D)-,
while $\gamma \rightarrow \infty $ implies 2D-behavior. The
resulting competition between anisotropy and superconductivity
raises serious doubts whether 2D mechanisms and models,
corresponding to the limit $\gamma =\infty $, can explain the
essential observations of superconductivity in the
cuprates\cite{tsphysB}.
There is mounting evidence that close to the phase transition line $%
T_{c}\left( p\right) $ thermal fluctuations dominate, while
quantum fluctuations dominate both, the QSI and QSN transitions.
Furthermore, due to the 3D to 2D crossover, tuned by the rise of
$\gamma $ with reduced dopant concentration, these fluctuations
are enhanced. For these reasons, mean-field treatments, including
the BCS theory are expected to apply far from the critical line
only. This singles out the low temperature region around optimum
doping $p_{m}$.

In the mean-field approximation the London penetration depth of an
anisotropic superconductor in the Meissner state is given by\cite
{chandra,hirschfeld}
\begin{equation}
\frac{1}{\lambda _{i}^{2}\left( T\right) }\cong \frac{e^{2}}{\pi ^{2}\hbar
c^{2}}\oint dS_{F}\frac{\text{v}_{F_{i}}\text{v}_{F_{i}}}{\left| \text{{\bf v%
}}_{F}\right| }\left( 1+2\int_{\Delta _{k}}^{\infty }dE_{k}\frac{\partial
f\left( E_{k}\right) }{\partial E_{k}}\frac{E_{k}}{\sqrt{E_{k}-\Delta
_{k}^{2}}}\right) .  \label{eq1}
\end{equation}
$dS_{F}$,v$_{F_{i}}$ and $\left| {\bf v}_{\text{F}}\right| $ are
respectively the surface element of the Fermi surface, Fermi
velocity in direction i and the magnitude of the Fermi velocity.
$f\left( E_{k}\right) $ is the Fermi function and $\Delta _{k}$
the energy gap in direction $k$. The i index refers to the
principal crystallographic directions a, b and c. The second term
which is negative, describes the decrease of $1/\lambda
_{i}^{2}\left( T\right) $ caused by the thermal population of
Bogoliubov quasi-particle levels with energy\ $E_{k}$, and it is
this quantity where the anisotropy and magnitude of the energy gap
enters. In this approximation $1/\lambda _{i}^{2}\left( T\right) $
vanishes close to $T_{c}$ as $1/\lambda _{i}^{2}\left( T\right)
=1/\lambda _{i0}^{2}\left( 1-T/T_{c}\right) ^{\nu }$ with $\nu
=1/2$, while their is mounting evidence for $\nu \simeq 2/3$ in
the experimentally accessible regime. Thus close to $T_{c}$ where
thermal fluctuations dominate the mean field treatment fails.
Otherwise, e.g. at sufficiently low temperatures and far away from
the QSI and QSN transitions the neglect of fluctuations appears to
be justified. Here Eq.(\ref{eq1}) reduces for noninteracting
quasiparticle excitations around the four d-wave nodes, which
dominate the leading low temperature behavior, to\cite{durst}
\begin{equation}
\frac{1}{\lambda _{ab}^{2}\left( T\right) }=\frac{1}{\lambda _{ab}^{2}\left(
0\right) }\left( 1-AT\right) ,\ A=\lambda _{ab}^{2}\left( 0\right) \ln 2%
\frac{k_{B}e^{2}\text{v}_{F}}{\hbar ^{2}c^{2}d\ \text{v}_{2}},\ \frac{1}{%
\lambda _{i}^{2}\left( 0\right) }\cong \frac{e^{2}}{\pi ^{2}\hbar c^{2}}%
\oint dS_{F}\frac{\text{v}_{F_{i}}\text{v}_{F_{i}}}{\left| \text{{\bf v}}%
_{F}\right| }.  \label{eq2}
\end{equation}
The Fermi velocities v$_{F}$ and v$_{2}$ enter the quasiparticles excitation
energies $E_{k}=\hbar \sqrt{\text{v}_{F}^{2}k_{1}^{2}+\text{v}%
_{2}^{2}k_{2}^{2}}$ and refer to velocities along directions normal and
tangential to the Fermi surface at each node. $d$ is the mean interlayer
spacing along the c-axis. The velocity ratio v$_{F}/$v$_{2}$ is a
fundamental material parameter which measures the anisotropy of the
quasiparticle excitation spectrum. This scenario is not restricted to the
penetration depth. It predicts simple power-law temperature dependencies in
the thermodynamic and transport properties at sufficiently low temperatures.
For example, the penetration depth measurements find that $1/\lambda
_{ab}^{2}$ exhibits in the clean limit and at low temperature a linear
temperature dependence\cite{hardy,tsjsup}, in agreement with Eq.(\ref{eq2}).
The NMR relaxation rate exhibits the expected $T^{3}$ temperature dependence
\cite{martindale}. The predicted effect of impurities in giving rise to a
universal thermal conductivity\cite{lee,graf} has been confirmed\cite
{taillefer}. The clean-limit specific heat varying as $T^{2}$ appears to
have been observed\cite{moler,wright,ywang}.

For a spherical Fermi surface, one recovers for the zero temperature
penetration depth the standard result, $1/\lambda ^{2}\left( T=0\right)
=4\pi ne^{2}/\left( mc^{2}\right) $, where $n$ is the number density of the
electrons in the normal state\cite{tinkham}. In recent years the penetration
depth has been the subject of intense experimental investigation in high-$%
T_{c}$ compounds \cite
{zhaobi,hofer,khasanovpr,khasanovd,tallonqcp,uemura,dicastro}. It has become
the practice to associate the anisotropy with anisotropic effective masses $%
m_{i}^{\ast }$ and to interpret the experimental data in terms of the London
formula
\begin{equation}
\frac{1}{\lambda _{i}^{2}\left( 0\right) }=\frac{4\pi n_{s}e^{2}}{%
m_{i}^{\ast }c^{2}}  \label{eq3}
\end{equation}
by introducing the number density $n_{s}$ of the superfluid. In a real
superconductor, the ionic potential modifies the spherical Fermi surface of
free electrons drastically. In this case it is not evident what should be
taken for the effective mass $m_{i}^{\ast }$ and the number density $n_{s}$.
Although, using Eqs.(\ref{eq2}) and (\ref{eq3}), we can define the ratio $%
n_{s}/m_{i}^{\ast }$, which often has been used to interpret experimental
results, in terms of

\begin{equation}
\ \frac{1}{\lambda _{i}^{2}\left( 0\right) }\cong \frac{e^{2}}{\pi ^{2}\hbar
c^{2}}\oint dS_{F}\frac{\text{v}_{F_{i}}\text{v}_{F_{i}}}{\left| \text{{\bf v%
}}_{F}\right| }=\frac{4\pi n_{s}e^{2}}{m_{i}^{\ast }c^{2}}\frac{n_{s}}{%
m_{i}^{\ast }}.  \label{eq4}
\end{equation}
This relation shows that $n_{s}/m_{i}$ is just a way of parameterizing
experimental results, with no discernible connection to the band mass or
carrier concentration. Indeed in the mean-field approximation $1/\lambda
_{i}^{2}\left( 0\right) $ is determined by normal state properties, namely
the integral of the Fermi velocity over the Fermi surface. Noting that in
high-$T_{c}$ superconductors band structure calculations\cite{picket} and
ARPES\cite{campuzano,lanzara} studies uncovered drastic deviations from the
free electron picture it is evident that interpretations based on the ratio $%
n_{s}/m_{i}^{\ast }$ obscure the origin of the anisotropy $\gamma
_{ij}=\lambda i/\lambda _{j}$ and the doping dependence of the zero
temperature penetration depth, as well as the isotope effect on this
quantity. Indeed, an inspection of Eq.(\ref{eq2}) leads to the conclusion
that the anisotropy stems from flat portions of the Fermi surface, while the
doping dependence reflects that of the Fermi surface. On the other hand ,
there is little doubt about the importance of residual electron-electron and
electron-phonon interactions, not accounted for in Eq.(\ref{eq2}).
Quantifying these interactions is difficult in the normal state of the
cuprates, given the lack of well-defined single-particle excitations as
revealed by various experiments. Contrariwise, well-defined quasiparticle
excitations do exist in the superconducting state, and a description of the
low temperature state in terms of superfluid Fermi liquid theory is believed
to apply. Fermi-liquid corrections account for the Fermi-liquid interactions
between electrons. In the superconductor their effect is the renormalization
of the Fermi velocity ratio in terms of v$_{F}/ $v$_{2}\rightarrow \alpha
^{2}$v$_{F}/$v$_{2}$\cite{durst}, where v$_{F}/$v$_{2}$ is the bare value
entering Eq.(\ref{eq2}). The comparison of $d\lambda ^{-2}/dT$ at $T=0$ in
Bi2212, evaluated with Eq.(\ref{eq2}) and the ARPES estimate for v$_{F}/$v$%
_{2}$\cite{campuzano} with the value deduced from penetration depth
measurements\cite{sflee} points to a substantial Fermi liquid
renormalization, namely $\alpha ^{2}\approx 0.3$, due to interactions
between the nodal quasiparticles in the superconducting state. To explore
the doping dependence of the renormalization we invoke the empirical
relation $T_{c}\left. d/dT\left( \lambda _{ab}^{2}\left( 0\right) /\lambda
_{ab}^{2}\left( T\right) \right) \right| _{T=0}\approx -0.6$\cite{tsjsup},
which applies to variety of cuprates with $T_{c}$ ranging from $30$ to $130$%
K and dopant concentrations extending from the underdoped to the optimally
doped regime. It implies with the empirical relation, $T_{c}\propto
1//\lambda _{ab}^{2}\left( 0\right) $\cite{uemura}, established for
underdoped cuprates, and Eq.(\ref{eq2}) that $\alpha ^{2}$v$_{F}/$v$_{2}$ is
nearly doping independent.

Much less attention has been devoted to the renormalization of the Fermi
velocity due to electron-phonon interaction. Isotope substitution, i.e. the
exchange of $^{18}$O by $^{16}$O is a suitable probe, whereby the lattice
parameters remain essentially unaffected\cite{conderla,raffa}, while the
phonon frequencies associated with the mass of the oxygen ions, or more
generally, the oxygen lattice degrees of freedom are modified\cite{wang}. In
recent years the isotope effect on the zero temperature penetration depth
has been investigated, using a variety of techniques. In Table I we listed
the experimental estimates for the relative change
\begin{equation}
\frac{\Delta \lambda _{ab}^{2}\left( \widetilde{T}\right) }{\lambda
_{ab}^{2}\left( \widetilde{T}\right) }=\frac{^{n}\lambda _{ab}^{2}\left(
\widetilde{T}\right) -^{m}\lambda _{ab}^{2}\left( \widetilde{T}\right) }{%
^{m}\lambda _{ab}^{2}\left( \widetilde{T}\right) },  \label{eq5}
\end{equation}
upon isotope exchange for various cuprates and MgB$_{2}$, where $n=^{18}$O, $%
m=^{16}$O in the cuprates and $n=^{11}$B, $m=^{10}$B in MgB$_{2}$. The data
also reveals that within experimental accuracy $\Delta \lambda
_{ab}^{2}\left( \widetilde{T}\right) /\lambda _{ab}^{2}\left( \widetilde{T}%
\right) =\Delta \lambda _{ab}^{2}\left( 0\right) /\lambda _{ab}^{2}\left(
0\right) $. At zero temperature and taking the anisotropy of the quoted
materials into account ($\lambda _{c}>>\lambda _{a}\approx \lambda _{b}$)
the mean-field expression (\ref{eq2}) reduces to
\begin{equation}
\frac{1}{\lambda _{ab}^{2}\left( T=0\right) }\cong \frac{e^{2}}{\pi
^{2}\hbar c^{2}\sqrt{2}}\oint dS\ \text{v}_{F_{ab}}.  \label{eq6}
\end{equation}
Since the lattice parameters remain essentially unaffected\cite
{conderla,raffa} by isotope exchange, while the dynamics, associated with
the mass of the respective ions, are modified, the substantial isotope
effect on the zero temperature penetration depth requires a renormalization
of the normal state Fermi velocity {\bf v}$_{F}\ \rightarrow \ \widetilde{%
{\bf v}}_{F}=\ ${\bf v}$_{F}/(1+f)$ where {\bf v}$_{F}$ is the bare velocity
and $f$ the electron-phonon coupling constant which changes upon oxygen
isotope exchange in the cuprates but remains nearly unaffected by boron
isotope substitution in MgB$_{2}$. However, in the Migdal-Eliashberg \cite
{me} (ME) treatment of the electron-phonon interaction the coupling constant
$f$ is independent of the ionic masses and assumed to be small \cite
{maksimov,carbotte}. This is true if the parameter $\omega _{0}f/E_{F}\ $is
small, where $\omega _{0}$ is the relevant phonon frequency and $E_{F}$ the
Fermi energy. Thus the isotope effect on the penetration depth is expected
to be small, of the order of the adiabatic parameter $\widetilde{\gamma }%
=\omega _{0}/E_{F}<<1$. The ME theory retains terms only of order 0.
Cuprates, however, have Fermi energies much smaller than those of
conventional metals \cite{randeria} so that $\widetilde{\gamma }$ is no
longer negligible small. Thus the large oxygen isotope effects on the zero
temperature in-plane penetration depth in the cuprates, listed in Table I,
poses a fundamental challenge to this understanding and calls for a theory
that goes beyond ME\cite{deppeler,deppeler2}.

\begin{center}
\bigskip
\begin{tabular}{|c|c|c|c|c|}
\hline
& $T_{c}$(K) & $\widetilde{T}$ & $\Delta \lambda _{ab}^{2}\left( \widetilde{T%
}\right) /\lambda _{ab}^{2}\left( \widetilde{T}\right) $ & Ref \\ \hline
YBa$_{2}$Cu$_{3}$O$_{7-\delta }$ & 89.0 & 10 & 0.05 & \cite{khasanovpr} \\
\hline
Y$_{0.7}\Pr_{0.3}$Ba$_{2}$Cu$_{3}$O$_{7-\delta }$ & 60.6 & 5 & 0.13 &  \\
\hline
Y$_{0.6}\Pr_{0.4}$Ba$_{2}$Cu$_{3}$O$_{7-\delta }$ & 45.3 & 5 & 0.11 &  \\
\hline
YBa$_{2}$Cu$_{3}$O$_{7-\delta }$ (film) & 89.3 & 4 & 0.05 & \cite{khasanovd}
\\ \hline
La$_{2-x}$Sr$_{x}$CuO$_{4+\delta },x=0.08$ & 19.5 & 0 & 0.10(2) & \cite
{hofer} \\ \hline
La$_{2-x}$Sr$_{x}$CuO$_{4+\delta },x=0.086$ & 22.4 & 0 & 0.08(1) &  \\ \hline
Bi$_{1.6}$Pb$_{0.4}$Sr$_{2}$Ca$_{2}$Cu$_{3}$O$_{10+\delta }$ & 107 & 0 & 0.05
& \cite{zhaobi} \\ \hline
MgB$_{2}$ & 38.5 & 0 & 0.02(2) & \cite{dicastro} \\ \hline
\end{tabular}
\end{center}

Table I: Experimental estimates for $\Delta \lambda _{ab}^{2}\left(
\widetilde{T}\right) /\lambda _{ab}^{2}\left( \widetilde{T}\right) =\left(
^{n}\lambda _{ab}^{2}\left( \widetilde{T}\right) -^{m}\lambda
_{ab}^{2}\left( \widetilde{T}\right) \right) /^{m}\lambda _{ab}^{2}\left(
\widetilde{T}\right) $ with $n=^{18}$O, $m=^{16}$O in the cuprates and $%
n=^{11}$B, $m=^{10}$B in MgB$_{2}$

\bigskip

Indeed, the relative shifts $\Delta \lambda _{ab}^{2}\left( \widetilde{T}%
\right) /\lambda _{ab}^{2}\left( \widetilde{T}\right) $ are substantial and
surprisingly close to $\left( ^{18}M_{o}-^{16}M_{o}\right) /^{16}M_{o}=0.125$%
. This differs fundamentally from the behavior of optic O - phonon
frequencies. The expected behavior, $\omega \propto M^{-1/2}$, was confirmed
in YBa$_{2}$Cu$_{3}$O$_{7-\delta }$ by measuring the frequency shift of the
transverse optic phonons (copper-oxygen stretching modes), yielding $\Delta
\omega /\omega \approx -0.06$, in agreement with $\Delta \omega /\omega
\approx $ $\left( ^{16}M_{o}/^{18}M_{o}\right) ^{1/2}-1=-0.057$ \cite{wang}.
In any case the observed oxygen isotope effect on the zero temperature
penetration depth uncovers together with Eq.(\ref{eq6}), a substantial
renormalization of the normal state Fermi velocity due to oxygen lattice
degrees of freedom, while in MgB$_{2}$ this renormalization due to boron
isotope exchange turns out to be marginal. This renormalization is also
expected to affect the superconducting properties. Taking the empirical
relation $T_{c}\left. d/dT\left( \lambda _{ab}^{2}\left( 0\right) /\lambda
_{ab}^{2}\left( T\right) \right) \right| _{T=0}\approx -0.6$\cite{tsjsup}
for granted, Eq.(\ref{eq2}) implies that
\begin{equation}
\frac{\Delta T_{c}}{T_{c}}=-\frac{\Delta A}{A}=-\frac{\Delta \lambda
_{ab}^{2}\left( 0\right) }{\lambda _{ab}^{2}\left( 0\right) }-\frac{\Delta
\left( \widetilde{\text{v}_{F}/\text{v}_{2}}\right) }{\left( \widetilde{%
v_{F}/v_{2}}\right) },  \label{eq7}
\end{equation}
where $\widetilde{\text{v}_{F}/\text{v}_{2}}$ is the bare ratio,
renormalized with respect to electron-phonon coupling. Noting that close to
optimum doping $\Delta T_{c}/T_{c}$ is negligible small, the oxygen isotope
exchange uncovers a substantial electron-phonon renormalization of \ v$_{F}/$%
v$_{2}$, characterizing the quasiparticles in the superconducting state.
Indeed, close to optimum doping $\Delta \left( \widetilde{\text{v}_{F}/\text{%
v}_{2}}\right) /\left( \widetilde{\text{v}_{F}/\text{v}_{2}}\right) \cong -$
$\Delta \lambda _{ab}^{2}\left( 0\right) /\lambda _{ab}^{2}\left( 0\right) $
holds and this quantity varies from 0.05 to 0.11 (see Table I). This effect,
providing direct evidence of the relevance of electron-phonon coupling in
the superconducting state, should be observable with ARPES. Noting that
thermally excited quasiparticles destroy superconductivity by driving $%
1/\lambda _{ab}^{2}\left( T\right) $ to zero, we can estimate $T_{c} $ by
extrapolating Eq.(\ref{eq2}) to $1/\lambda _{ab}^{2}\left( T\right) =0$.
This yields $T_{c}A\approx 1$ and confirms Eq.(\ref{eq7}). In this context
it is interesting to note that details of the Fermi-surface topology of
deuterated $\kappa $-(BEDTTTF)$_{2}$Cu(NCS)$_{2}$ have been measured as a
function of pressure and compared with equivalent measurements of the
undeuterated salt. The data suggest that the negative isotope effect
observed on deuteration is due to small differences in Fermi-surface
topology caused by the isotopic substitution\cite{biggs}.

Close to the phase transition line $T_{c}\left( p\right) $ thermal critical
fluctuations, neglected in mean-field treatments, dominate the thermodynamic
properties. Approaching the phase transition line around $p=p_{m}$ from
below (see Fig.\ref{fig1}), there is mounting evidence that the critical
behavior of homogeneous cuprates falls in the experimentally accessible
temperature regime into the 3D-XY universality class\cite
{tsphysB,ffh,tshkws,ohl,hub,kamal,kamal2,pasler,tsjh,tsjs,jhts,tsjshou,book,meingast,osborn}%
. Here critical 3D-XY fluctuations dominate because the fluctuations of the
vector potential are strongly suppressed due to the small value of the
effective charge of the pairs\cite{ffh}. In the 3D-XY universality class the
transition temperature $T_{c}$, the critical amplitude of the specific heat $%
A^{-}$ and the critical amplitudes of the penetration depths $\lambda _{i0}$
are universally related by\cite{book}
\begin{equation}
\left( k_{B}T_{c}\right) ^{3}=\left( \frac{\Phi _{0}^{2}R^{-}}{16\pi ^{3}}%
\right) ^{3}\frac{1}{\lambda _{a0}^{2}\lambda _{b0}^{2}\lambda _{c0}^{2}A^{-}%
} .  \label{eq8}
\end{equation}
$R^{-}\simeq 0.815$ is a universal number. The critical amplitudes are
defined as $c=\left( A^{-}/\alpha \right) t^{-\alpha }$ , $t=1-T/T_{c}$ and $%
\lambda _{i}^{2}=\lambda _{i0}^{2}t^{-\nu }$, where $\alpha $ and
$\nu $ are the critical exponents. Although $T_{c}$, $A^{-}$ and
$\lambda _{i0}^{2}$ depend on the dopant concentration, isotope
exchange etc., universality implies that this combination does
not. Hence this relation puts a crucial constraint on the
microscopic theory of superconductivity in cuprates. To illustrate
this point it is instructive to consider the doping dependence of
$T_{c}$, $\lambda _{ab}^{2}\simeq \lambda _{a0}\lambda _{b0}$,
$\gamma _{0}=\lambda _{c0}/\lambda _{ab0}$ and $A^{-}$ close to
the 2D-QSI
transition (see Fig.\ref{fig1}). Here $T_{c}$, $1/\lambda _{ab0}^{2}$ and $%
\gamma _{0}$ are known to scale as $T_{c}\propto 1/\lambda _{ab0}^{2}\propto
1/\gamma _{0}\propto p-p_{u}$ so that the critical amplitude of the specific
heat singularity vanishes according to Eq.(\ref{eq8}) as $A^{-}\propto
\left( p-p_{u}\right) ^{2}\propto T_{c}^{2}$. This is consistent with the
specific heat data for underdoped La$_{2-x}$Sr$_{x}$CuO$_{4}$ and underdoped
Tl$_{2}$Ba$_{2}$CuO$_{6+\delta }$\cite{loram}. Furthermore, the universal
relation implies that the relative changes upon isotope substitution satisfy
the relation\cite{tstclaa}
\begin{equation}
\frac{\Delta T_{c}}{T_{c}}=-\frac{\Delta A^{-}}{3A^{-}}-\frac{1}{3}%
\sum_{i=a,b,c}\frac{\Delta \lambda _{i0}^{2}}{\lambda _{i0}^{2}}.
\label{eq9}
\end{equation}
It explains why close to the transition temperature and optimum doping ($%
p=p_{m}$) a substantial isotope effect on the penetration depths is
compatible with a negligible effect on $T_{c}$.

Approaching the 2D-XY-QSI transition in the underdoped regime (see Fig.\ref
{fig1}) a crossover to the universal relation\cite{book}

\begin{equation}
T_{c}=\frac{\Phi _{0}^{2}\overline{R}_{2}}{16\pi ^{3}k_{B}}\frac{d_{s}}{%
\lambda _{ab}^{2}\left( 0\right) },  \label{eq10}
\end{equation}
takes place and the cuprates correspond to a stack of independent sheets of
thickness $d_{s}$. $\overline{R}_{2}$ is a universal number. The flow to
2D--XY-QSI behavior is experimentally well confirmed in terms of Uemura's
plot\cite{uemura}. It is a characteristic 2D property and also applies to
the onset of superfluidity in $^{4}$He films adsorbed on disordered
substrates where it is well confirmed\cite{crowell}. Although $T_{c}$, $%
d_{s} $ and $\lambda _{ab}^{2}\left( 0\right) $ depend on dopant
concentration, isotope substitution, etc., universality implies
that this relation does not. This puts yet another constraint on
the microscopic theory of superconductivity in the cuprates.
Furthermore it yields for the relative changes upon isotope
exchange the relation
\begin{equation}
\frac{\Delta T_{c}}{T_{c}}=\frac{\Delta d_{s}}{d_{s}}-\frac{\Delta \lambda
_{0ab}^{2}\left( 0\right) }{\lambda _{ab}^{2}\left( 0\right) },  \label{eq11}
\end{equation}
applicable close to the underdoped limit ($p=p_{u}$). Although the
experimental data are rather sparse in the underdoped regime\cite
{tstclaa,tshk}, suggestive evidence for an isotope effect on the
effective thickness $d_{s}$ of the superconducting sheet emerges,
namely $\Delta d_{s}/d_{s}=\left( ^{18}d_{s}-^{16}d_{s}\right)
/^{16}d_{s}\approx 0.03$\cite {tstclaa}. Estimates for $d_{s}$ can
be derived from the crossing point phenomenon in the temperature
dependence of the magnetization for various applied magnetic
fields, applied parallel to the c-axis. In a 3D anisotropic
superconductor, falling into the 3D-XY universality class, the
magnetization data plotted in terms $m_{z}/H_{z}^{1/2}$ versus $T$
will cross at $T_{c}$, where $z$ is along the c-axis. Consistency
with this behavior was found in a variety of cuprates\cite{book}.
On the contrary, in a 2D superconductor, corresponding to a slab
of thickness $d_{s}$, the crossing point occurs in the plot
$m_{z}$ versus $T$ at the Kosterlitz-Thouless transition
temperature $T_{KT}$, where $m_{z}\propto -k_{B}T_{KT}/\left( \Phi
_{0}d_{s}\right) $. Even though bulk cuprates are strictly 2D only
close to the 2D-QSI transition at $p=p_{u}$ (see Fig.\ref{fig1}),
in the highly anisotropic materials, such as Bi-2212 and Tl-1223,
2D crossing point feature have been observed and used to estimate
$d_{s}$, yielding values close to the c-axis lattice
constant\cite{book}. Since the lattice parameters remain
essentially unaffected\cite{conderla,raffa} by isotope exchange,
while $d_{s}$ does, the substantial relative change $\Delta
d_{s}/d_{s}=\left( ^{18}d_{s}-^{16}d_{s}\right) /^{16}d_{s}\approx
0.03$\cite {tstclaa} uncovers again the relevance of
electron-lattice coupling. Going further, by combining
Eqs.(\ref{eq7}) and (\ref{eq11}), extrapolated to the underdoped
regime, we obtain the approximate relation
\begin{equation}
\frac{\Delta d_{s}}{d_{s}}=-\frac{\Delta \left( \widetilde{\text{v}_{F}/%
\text{v}_{2}}\right) }{\left(
\widetilde{\text{v}_{F}/\text{v}_{2}}\right)}, \label{eq11a}
\end{equation}
which provides additional evidence for the coupling between
superconducting properties and lattice degrees of freedom. The
approximate nature of this relation stems from the fact that the
d-wave quasiparticle scenario does not hold down to the 2D-QSI
transition due to the neglect of quantum fluctuations, associated
with the phase of the order parameter. Indeed, the linear-in-T
temperature dependence of $1/\lambda _{ab}^{2}$ simply follows
from the existence of a 2D-QSI transition. The result is, $\left(
\lambda _{ab}\left( 0\right) /\lambda _{ab}\left( T\right) \right)
^{2}-1\propto T/T_{c}$, holds for d- and s-wave
pairing\cite{tsjsup}.

\begin{figure}[tbp]
\centering
\includegraphics[totalheight=6cm]{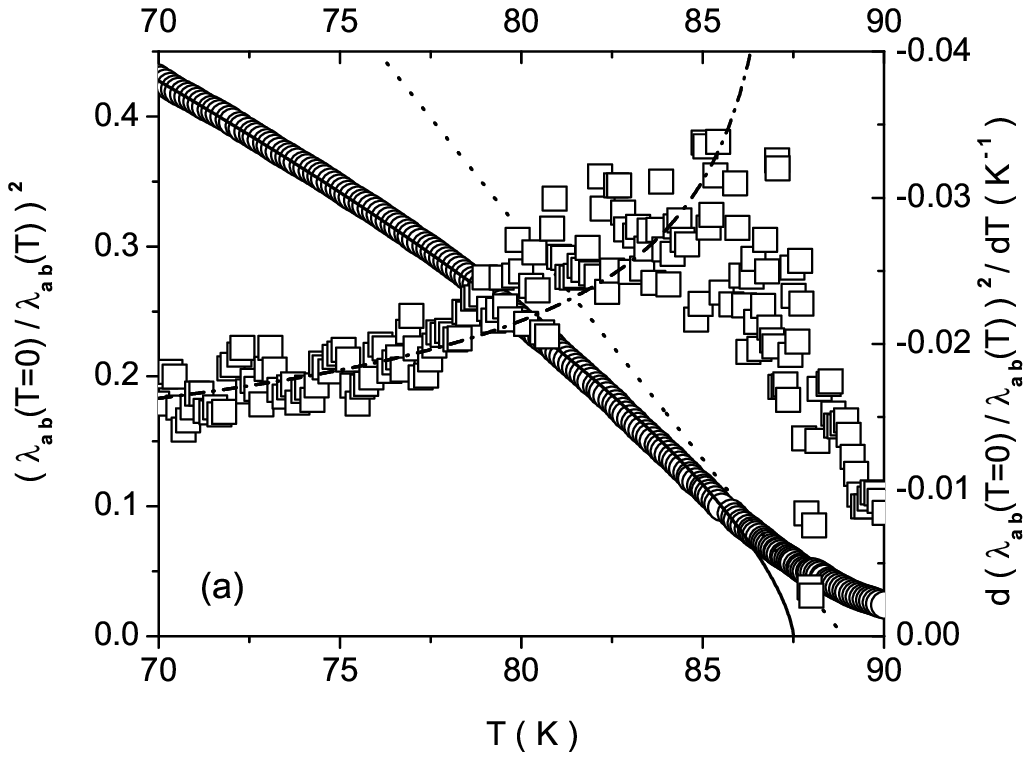}
\includegraphics[totalheight=6cm]{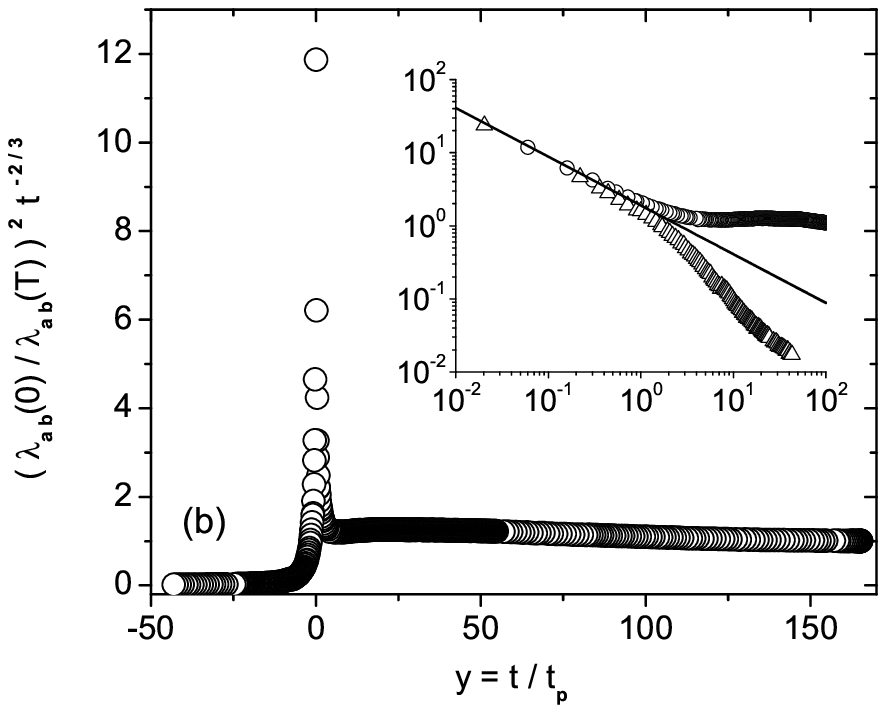}
\caption{(a)Microwave surface impedance data for $\protect\lambda
_{ab}^{2}\left( T=0\right) /\protect\lambda _{ab}^{2}\left(
T\right) $ ($\bigcirc $)\ and $d\left( \protect\lambda
_{ab}^{2}\left( T=0\right) /\protect\lambda _{ab}^{2}\left(
T\right) \right) /dT$ ($\square $) versus $T$ of a high-quality
Bi$_{2}$Sr$_{2}$CaCu$_{2}$O$_{8+\protect\delta }$ single crystal
taken from Jacobs {\em et al.} \protect\cite{jacobs}. The solid
line is $\protect\lambda _{ab}^{2}\left( 0\right) /\protect\lambda
_{ab}^{2}\left( T\right) =1.2\left( 1-T/T_{c}\right) ^{2/3}$ and
the dash-dot line its derivative with $T_{c}=87.5$K, indicating
the leading critical behavior of the homogeneous system. The
dotted line is the tangent to the inflection point at
$T_{p}\approx 87$K, where $d\left( \protect\lambda _{ab}^{2}\left(
0\right) /\protect\lambda _{ab}^{2}\left( T\right) \right) /dT$ is
maximum; (b) Finite size scaling function $g\left( y\right)
=\left( \protect\lambda _{0ab}/\protect\lambda _{ab}\left(
T\right) \right) ^{2}\left| t\right| ^{-\protect\nu }$ versus
$y=t/\left| t_{p}\right| $ for the data shown in Fig.\ref{fig2}a.
The solid line in the inset is Eq.(\ref {eq18}) with
$g_{0c}=1.6$.} \label{fig2}
\end{figure}

Additional evidence for electron-lattice coupling emerges from the combined
oxygen isotope and finite size effects. Due to inhomogeneities of extrinsic
or intrinsic origin, cuprates are homogeneous over a finite extent only.
Thus a finite size effect\cite{fsize} is expected to occur, whereby the
correlation volume cannot grow beyond the volume of the homogeneous domains.
When 3D-XY critical fluctuations dominate there is the universal relationship
\cite{book}
\begin{equation}
\frac{1}{\lambda _{i}\left( T\right) \lambda _{j}\left( T\right) }=\frac{%
16\pi ^{3}k_{B}T}{\Phi _{0}^{2}\sqrt{\xi _{i}^{t}\left( T\right) \xi
_{j}^{t}\left( T\right) }},  \label{eq12}
\end{equation}
between the London penetration depths $\lambda _{i}$ and transverse
correlation lengths $\xi _{i}^{t}$ in directions $i$ and $j$. In the
presence of inhomogeneities with length scales $L_{i}$ the $\xi _{i}^{t}$ $%
=\xi _{i0}^{t}\left| t\right| ^{-\nu }$, where $t=T/T_{c}-1$, cannot diverge
but are bounded by
\begin{equation}
\xi _{i}^{t}\xi _{j}^{t}\leq L_{k}^{2},\ i\neq j\neq k.  \label{eq13}
\end{equation}
A characteristic feature of the resulting finite size effect is the
occurrence of an inflection point at $T_{p_{k}}$ in $1/\lambda _{i}\left(
T\right) \lambda _{j}\left( T\right) $ below $T_{c}$, the transition
temperature of the homogeneous system. Here
\begin{equation}
\xi _{i}^{t}\left( T_{p_{k}}\right) \xi _{j}^{t}\left( T_{p_{k}}\right)
=L_{k}^{2},\ i\neq j\neq k,  \label{eq14}
\end{equation}
and Eq.(\ref{eq12}) reduces to
\begin{equation}
\left. \frac{1}{\lambda _{i}\left( T\right) \lambda _{j}\left( T\right) }%
\right| _{T=T_{p_{k}}}=\frac{16\pi ^{3}k_{B}T_{p_{k}}}{\Phi _{0}^{2}}\frac{1%
}{L_{k}}.  \label{eq15}
\end{equation}
In the homogeneous case $1/\left( \lambda _{i}\left( T\right) \lambda
_{j}\left( T\right) \right) $ decreases continuously with increasing
temperature and vanishes at $T_{c}$, while in the presence of
inhomogeneities it exhibits an inflection point at $T_{p_{k}}<T_{c}$, so
that
\begin{equation}
\left. d\left( \frac{1}{\lambda _{i}\left( T\right) \lambda
_{j}\left( T\right) }\right) /dT\right| _{T=T_{p_{k}}}=0
\label{eq16}
\end{equation}
Since the experimental data for the temperature dependence of the
penetration depths is available in the form $\lambda _{ab}$ and
$\lambda _{c} $ only, we rewrite Eq.(\ref{eq15}) as
\begin{equation}
L_{c}=\frac{16\pi ^{3}k_{B}T_{p_{c}}\lambda _{ab}^{2}\left( T_{p_{c}}\right)
}{\Phi _{0}^{2}},\ L_{ab}=\frac{16\pi ^{3}k_{B}T_{p_{b}}\left( \lambda
_{ab}\left( T\right) \lambda _{c}\left( T\right) \right) _{T=T_{p_{ab}}}}{%
\Phi _{0}^{2}}.  \label{eq17}
\end{equation}
Apart from the inflection point, an essential characteristic of a
finite size effect is the finite size scaling
function\cite{schultka}. In the present case \ and for $\lambda
_{ab}$ it is defined in terms of
\begin{equation}
\left( \frac{\lambda _{0ab}}{\lambda _{ab}\left( T\right) }\right)
^{2}\left| t\right| ^{-\nu }=g_{c}\left( y\right) ,\ \
y=sign\left( t\right) \left| t\right| \left( \frac{L_{c}}{\xi
_{0ab}^{t}}\right) ^{1/\nu }=sign\left( t\right) \left|
\frac{t}{t_{p_{c}}}\right|   \label{eq18}
\end{equation}
For $t$ $=T/Tc-1$ small and $L_{c}\rightarrow \infty $, so that
$\pm y\rightarrow \infty $, it should tend to $g_{c}\left(
y\rightarrow -\infty \right) =1$ and $\ g_{c}\left( y\rightarrow
\infty \right) =0$, respectively, while for $t=0$ and $L_{c}\neq
0$ it diverges as
\begin{equation}
g_{c}\left( y\rightarrow 0\right) =g_{0c}\left| y\right| ^{-\nu
}=g_{0c}\left| \frac{t}{t_{p_{c}}}\right| ^{-\nu },  \label{eq19}
\end{equation}
whereby $\left( \lambda _{0ab}/\lambda _{ab}\left( T_{c},L\right)
\right) ^{2}=g_{0c}\left| t_{p_{c}}\right| ^{\nu }=g_{0c}\xi
_{0ab}^{t}/L_{c}$. As expected, a sharp superconductor to normal
state transition requires domains of infinite extent. Moreover at
$t_{p_{c}}$, $y=1$ and there is an inflection point because
$d\left( \lambda _{0ab}/\lambda _{ab}\left( T,L\right) \right)
^{2}/dt=0$. Since the scaling function $g_{c}\left( y\right) $
depends on the type of confining geometry and on the conditions
imposed (or not, in the case of free boundaries) at the boundaries
of the domains, this applies to the amplitude $g_{0c}$ as well. In
Fig.\ref{fig2}a we displayed the microwave surface impedance data
for $\lambda
_{ab}^{2}\left( T=0\right) /\lambda _{ab}^{2}\left( T\right) $ \ and $%
d\left( \lambda _{ab}^{2}\left( T=0\right) /\lambda
_{ab}^{2}\left( T\right)
\right) /dT$ versus $T$ of a high-quality Bi$_{2}$Sr$_{2}$CaCu$_{2}$O$%
_{8+\delta }$ single crystal taken from Jacobs {\em et al.}
\cite{jacobs}. The solid curve indicates the leading 3D-XY
critical behavior of the homogeneous system, while the data
uncovers a rounded transition which occurs smoothly. This
behavior, together with the occurrence of an inflection point
around $T_{p_{c}}\approx 87$K, where $d\left( \lambda
_{ab}^{2}\left( T=0\right) /\lambda _{ab}^{2}\left( T\right)
\right) /dT$ exhibits an extremum, points to a finite size effect.
With $\lambda _{ab}\left( T=0\right) =1800$\AA\ obtained from $\mu
$SR measurements \cite {leem}, $T_{p_{c}}\approx 87$K and $\lambda
_{ab}^{2}\left( T=0\right) /\lambda _{ab}^{2}\left(
T_{P_{c}}\right) =0.066$ we obtain with the aid of Eq.(\ref{eq17})
the estimate $L_{c}\approx 68$\AA . Although the spatial extent of
the homogeneous domains along the c-axis appears to be of
nanoscale only, the small critical amplitude of the transverse
correlation length, $\xi _{0ab}^{t}=L_{c}\left(
1-T_{p_{c}}/T_{c}\right) ^{2/3}\approx 2.2$\AA, makes the 3D-XY
critical regime ($\lambda _{ab}^{2}\left( 0\right) /\lambda
_{ab}^{2}\left( T\right) \propto \left( 1-T/T_{c}\right) ^{2/3}$)
attainable. An additional and essential characteristic of a finite
size effect appearing in the temperature dependence of the
in-plane penetration depth is the consistency of \ $\left( \lambda
_{0ab}/\lambda _{ab}\left( T\right) \right) ^{2}\left| t\right|
^{-\nu }$ versus $t/\left| t_{p_{c}}\right| $ with the shape and
limiting behavior of the finite size scaling function (see
Eq.(\ref{eq18})). In Fig.\ref{fig2}b we displayed $\left( \lambda
_{0ab}/\lambda _{ab}\left( T\right) \right) ^{2}\left| t\right|
^{-\nu }$ versus $t/\left| t_{p_{c}}\right| $. The apparent
agreement with the aforementioned characteristic behavior of this
function, provides strong evidence for a finite size effect, due
to the limited extent $L_{c}$ of homogeneous superconducting
domains along the c-axis. Clearly, such a finite size scaling
analysis, performed on one set of data for one particular sample
and material only, cannot distinguish between an intrinsic or
extrinsic origin of the inhomogeneity. Noting that this behavior
was found in a variety of cuprates and for data obtained with
different techniques\cite{bled}, one is lead to the conclusion
that inhomogeneities, giving rise to a finite size effect, are yet
another facet of the mystery of superconductivity in the cuprates.
Clearly this finite size effect is not restricted to the
penetration depth but should be visible in other thermodynamic
properties. In the specific heat it leads to a rounding of the
peak and its consistency with a finite size effect was established
for the data taken on YBa$_{2}$Cu$_{3}$O$_{7-\delta }$ high
quality single crystals\cite{book}. In these samples the domain
size was found to range from 300 to 400 \AA . Furthermore
nanoscale spatial variations in the electronic characteristics
have also been observed in underdoped
Bi$_{2}$Sr$_{2}$CaCu$_{2}$O$_{8+\delta }$ with scanning tunnelling
microscopy (STM)\cite{liu,chang,cren,lang}. They reveal a spatial
segregation of the electronic structure into 3nm diameter
superconducting domains in an electronically distinct background.

Supposing that the limiting length scales change upon isotope exchange, we
obtain from Eq.(\ref{eq17}) the relation
\begin{equation}
\frac{\Delta T_{p_{c}}}{T_{p_{c}}}=\frac{\Delta L_{c}}{L_{c}}-\frac{\Delta
\lambda _{ab}^{2}\left( T_{p_{c}}\right) }{\lambda _{ab}^{2}\left(
T_{p_{c}}\right) },  \label{eq20}
\end{equation}
which matches Eq.(\ref{eq11}), applicable in the 2D limit where the limiting
length is set by $d_{s}$, the thickness of the independent sheets.
Furthermore, this relation opens the possibility to probe the coupling
between superfluidity and lattice degrees of freedom close to criticality,
where mean-field treatments fail. Indeed, given the fact that the lattice
parameters remain essentially unaffected\cite{conderla,raffa} by isotope
exchange, a purely electronic mechanism requires $\Delta L_{c}=0$. The
effect of oxygen isotope substitution on the inhomogeneity induced finite
size effect has been explored in Y$_{1-x}$Pr$_{x}$Ba$_{2}$Cu$_{3}$O$%
_{7-\delta }$\cite{tsrkhk}. From the resulting estimates, listed in Table
II, several observations emerge. First, $L_{c}$ increases systematically
with reduced $T_{p_{c}}$. Second, $L_{c}$ grows with increasing $x$ and upon
isotope exchange ($^{16}$O, $^{18}$O). Third, the relative shift of $%
T_{p_{c}}$ is very small. This reflects the fact that the change of $L_{c}$
is essentially due to the superfluid, probed in terms of $1$/$\lambda
_{ab}^{2}$. Accordingly, $\Delta L_{c}/L_{c}\approx \Delta \lambda
_{ab}^{2}\left( T_{p_{c}}\right) /\lambda _{ab}^{2}\left( T_{p_{c}}\right) $
for $x=0,\ 0.2$ and $0.3$.

\bigskip

\begin{center}
\begin{tabular}{|c|c|c|c|}
\hline
x & 0 & 0.2 & 0.3 \\ \hline
$\Delta T_{p_{c}}/T_{p_{c}}$ & -0.000(2) & -0.015(3) & -0.021(5) \\ \hline
$\Delta L_{c}/L_{c}$ & 0.12(5) & 0.13(6) & 0.16(5) \\ \hline
$\Delta \lambda _{ab}^{2}\left( T_{p_{c}}\right) /\lambda _{ab}^{2}\left(
T_{p_{c}}\right) $ & 0.11(5) & 0.15(6) & 0.15(5) \\ \hline
$^{16}T_{p_{c}}(K)$ & 89.0(1) & 67.0(1) & 52.1(2) \\ \hline
$^{18}T_{p_{c}}\left( K\right) $ & 89.0(1) & 66.0(2) & 51.0(2) \\ \hline
$^{16}L_{c}\left( \text{\AA }\right) $ & 9.7(4) & 14.2(7) & 19.5(8) \\ \hline
$^{18}L_{c}\left( \text{\AA }\right) $ & 10.9(4) & 16.0(7) & 22.6(9) \\
\hline
\end{tabular}
\end{center}

\bigskip

Table II: Finite size estimates for$\Delta T_{p_{c}}/T_{p_{c}}$, $\Delta
L_{c}/L_{c}$ ,$\Delta \lambda _{ab}^{2}\left( T_{p_{c}}\right) /\lambda
_{ab}^{2}\left( T_{p_{c}}\right) $, $^{16}T_{p_{c}}$, $^{18}T_{p_{c}}$, $%
^{16}L_{c}$ and $^{18}L_{c}$ for an $^{\text{18}}$O content of 89\% taken
from \cite{tsrkhk}. \bigskip

To appreciate the implications of these estimates, we note again that for
fixed Pr concentration the lattice parameters remain essentially unaffected
\cite{conderla,raffa}. Accordingly, an electronic mechanism, without
coupling to lattice degrees of freedom implies $\Delta L_{c}=0$. On the
contrary, a significant change of $L_{c}$ upon oxygen exchange uncovers the
coupling between the superfluid, probed by $\lambda _{ab}^{2}$, and the
oxygen lattice degrees of freedom. A glance to Table II shows that the
relative change of the superconducting domains along the c-axis upon oxygen
isotope exchange is significant, ranging from $12$ to $16$\%, while the
relative change of the inflection point $T_{p_{c}}$ is an order of magnitude
smaller. For this reason the significant relative change of $L_{c}$ at fixed
Pr concentration is accompanied by essentially the same relative change of $%
\lambda _{ab}^{2}$, which probes the superfluid. This uncovers unambiguously
the existence and relevance of the coupling between the superfluid and
oxygen lattice degrees of freedom. Furthermore, this behavior agrees with
the isotope effect on $d_{s}$\cite{tstclaa}, the limiting length scale close
to the 2D-QSI transition. Potential candidates for the relevant lattice
degrees of freedom are the Cu-O bond-stretching-type phonons showing
temperature dependence, which parallels that of the superconductive order
parameter\cite{chung}.

An additional probe to unravel the mystery of superconductivity in
the cuprates is the response to a magnetic field. In the early
discussion of the symmetry of the order parameter, Yip and
Sauls\cite{yip} proposed that the angular position of the gap
nodes could be probed by a measurement of the magnetic field
dependence of the penetration depth. In the local limit and for
T$\rightarrow $0, they predicted the linear relationship, $\left(
\lambda \left( H=0,T=0\right) /\lambda \left( H,0\right) \right)
^{2}-1\propto -H$, where the factor of proportionality is
independent of temperature. Several experimental groups tried to
verify this prediction, but failed to identify a linear $H$ term
which scaled with temperature according to the
theory\cite{maeda,carrington,bidinosti,sonier}. On the other hand,
calculations based on a d-wave model, treated in the quasi
classical approximation, suggest that $\left( \lambda \left(
H=0,T=0\right) /\lambda \left( H,0\right) \right) ^{2}-1\propto
-\sqrt{H}$\cite{sharapov}. However, in these treatments
fluctuations have been neglected. Close to the phase transition
line $T_{c}\left( p\right) $, where thermal fluctuations dominate,
the in-plane penetration depth scales as $\lambda _{ab}^{2}\propto
\xi _{c}=\xi _{ab}/\gamma $, while the magnetic field applied
parallel to the c-axis scales as $H_{c}\propto \Phi _{0}/\xi
_{ab}^{2}$. Thus the in-plane penetration satisfies at $T_{c}$ the
scaling form
\begin{equation}
\frac{1}{\lambda _{ab}^{2}\left( T_{c},H_{c}\right) }\propto
\sqrt{H_{c}}, \label{eq21}
\end{equation}
revealing that a superconductor is dramatically influenced by an
applied magnetic field. This behavior can be understood by noting
that in an applied magnetic field the correlation length cannot
grow indefinitely. For nonzero magnetic field $H_{c}$ there is the
limiting length scale $L_{H_{c}}\simeq \sqrt{\Phi _{0}/\left(
aH_{c}\right) }$ with $a\simeq 3.12$\cite{bled}, related to the
average distance between vortex lines. Indeed, as the magnetic
field increases the density of vortex lines becomes greater, but
this cannot continue indefinitely. The limit is roughly set on the
proximity of vortex lines by the overlapping of their cores. Due
to this limiting length scale the phase transition of a
homogeneous superconductor is rounded and occurs smoothly. At
$T=0$ and close to the 2D-QSI transition the
in-plane penetration depth scales as $\lambda _{ab}^{2}\propto \xi _{ab}^{z}$%
, where $z$ is the dynamic critical exponent of this transition. With $%
H_{c}\propto \Phi _{0}/\xi _{ab}^{2}$ this yields the scaling form
\begin{equation}
\left( \frac{\lambda \left( 0,0\right) }{\lambda \left( H_{c},T=0\right) }%
\right) ^{2}-1\propto -H_{c}^{z/2}.  \label{eq22}
\end{equation}
Contrariwise at $T=0$ and close to the 3D-QSN transition the
in-plane penetration depth scales as $\lambda _{ab}^{2}\propto \xi
_{ab}^{z+1}$ so that
\begin{equation}
\left( \frac{\lambda \left( 0,0\right) }{\lambda \left( H_{c},T=0\right) }%
\right) ^{2}-1\propto -H_{c}^{\left( z+1\right) /2}.  \label{eq23}
\end{equation}
Since the order parameter is assumed to be a complex scalar, these
scaling forms hold for both, s-wave and d-wave pairing. Taking the
evidence for a 2D-QSI transition with $z=1$ and a 3D-QSN
transition with $z=2$ into account \cite{tsphysB}, the scaling
form $\left( \lambda \left( H=0,T=0\right) /\lambda \left(
H,0\right) \right) ^{2}-1\propto -\sqrt{H_{c}}$ is expected to
hold in the underdoped regime, while in the overdoped limit
$\left( \lambda \left( H=0,T=0\right) /\lambda \left( H,0\right)
\right) ^{2}-1\propto -H_{c}^{3/2}$ should apply. In any case more
extended experimental investigations are required, including
samples covering the full doping range, to overcome the present
impasse. Another property suite to shed light on the critical
properties of the quantum transitions is the magnetic field
dependence of the zero temperature specific heat coefficient. At
$T=0$ and close to the 2D-QSI or 3D-QSN transitions it scales as
$\left(
c/T\right) _{T=0}\propto \xi _{ab}^{D-z}\propto H_{c}^{\left( D-z\right) /2}$%
. The data taken on La$_{2-x}$Sr$_{x}$CuO$_{4}$\cite{chen} points
to $\left( D-z\right) /2\simeq 1/2$, irrespective of the dopant
concentration. This suggests respectively, $z=1$ for the 2D-QSI
and $z=2$ for the 3D-QSN transition.

\begin{figure}[tbp]
\centering
\includegraphics[totalheight=6cm]{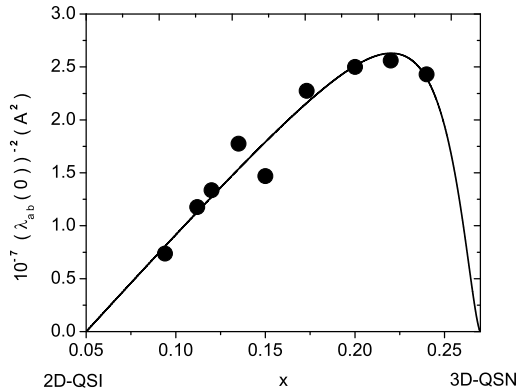}
\caption{$1/\lambda _{ab}^{2}\left( 0\right) $ versus $p\simeq x$
for La$_{2-x} $Sr$_{x}$CuO$_{4}$. $\bullet $: experimental data
taken from Panagopoulos {\em et al.}\protect\cite{panagop} The
solid line indicate the crossover from a 2D-QSI transition with
$z=\overline{\nu }=1$ to a 3D-QSN transition with $z=2 $ and
$\overline{\nu }=1/2$.} \label{fig3}
\end{figure}

The doping dependence of the zero temperature penetration depths
provides another link between the quantum critical behavior in the
underdoped and overdoped regimes. In Fig.\ref{fig3} we displayed
$1/\lambda _{ab}^{2}\left( 0\right) $ versus $p\simeq x$ for
La$_{2-x}$Sr$_{x}$CuO$_{4}$ taken from Panagopoulos {\em et
al.}\cite{panagop}. Close to the quantum phase transitions
$1/\lambda _{ab}^{2}\left( 0\right) $ scales as $1/\lambda
_{ab}^{2}\left( 0\right) \propto \delta ^{\overline{\nu }\left(
D+z-2\right) }$\cite{book}, where $\delta =p-p_{u}$ at the 2D-QSI
and $\delta =p_{o}-p$ at the 3D-QSN transition. The solid line
indicates the crossover from a 2D-QSI transition with
$z=\overline{\nu }=1$ to a 3D-QSN transition with $z=2 $ and
$\overline{\nu }=1/2$. \ While the flow to the 2D-QSI transition
is apparent, the data does not extend sufficiently close to 3D-QSN
criticality to confirm this crossover unambiguously. In any case,
it emerges that the properties of the ground state are controlled
by the crossover from the 2D-QSI to the 3D-QSN critical point. For
this reason it can be understood that away from these quantum
critical points, around optimum doping ($p=p_{m} $, see
Fig.\ref{fig1}), quantum fluctuations are suppressed to the extent
that Bogoliubov quasi-particle features are observable and
mean-field treatments represent a reasonable starting point. A
2D-QSI transition with $z=1$ and $\overline{\nu }=1$ coincides
with the theoretical prediction for a 2D disordered bosonic system
with long-range Coulomb interactions\cite{mpaf,cha,herbut}. A
potential candidate for 3D-QSN criticality is the disordered
d-wave superconductor to disordered metal transition at weak
coupling considered by Herbut\cite{herbutqsn} [38], with $z=2$ and
$\overline{\nu }=1/2$. energy. Here the disorder destroys
superconductivity, while at the 2D-QSI transition it localizes the
pairs.

To summarize, in the regime where mean-field treatments are
expected to apply, the substantial isotope effect on the zero
temperature penetration depth, established by a variety of
experimental techniques, implies a renormalization of the normal
state Fermi velocity due to a electron-lattice coupling, beyond
the band velocity and the ME theory. This coupling also affects
the superconducting properties and should be observable with
ARPES. Close to the critical line $T_{c}\left( p\right) $ and the
critical endpoints, either thermal or quantum fluctuations, not
included in mean-field treatments, dominate. In these regions of
the phase diagram the theory of quantum and thermal critical
phenomena applies. Given the universality class of the respective
transition, there are universal relations between critical
properties, putting stringent constraints on the microscopic
theory of superconductivity in the cuprates. Along the phase
transition line there is in the experimentally accessible
temperature regime mounting evidence for 3D-XY universality and
for a 3D-2D- crossover as the underdoped limit is approached. Here
a 2D-QSI transition occurs. This crossover, measured in terms of
the ratio $\gamma =\lambda _{c}/\lambda _{ab} $ (see
Fig.\ref{fig1}) is well documented\cite{tseuro,tsphysB} and
implies that superconductivity in the cuprates is a genuine 3D
phenomenon. The resulting competition between anisotropy and
superconductivity raises serious doubts whether 2D mechanisms and
models, corresponding to the limit $\gamma =\infty $, can explain
the essential observations of superconductivity in the cuprates.
Indeed, as the dopant concentration is increased, the cuprates
undergo in the ground state a crossover from 2D-QSI to 3D-QSN
criticality. For this reason the observation of  Bogoliubov
quasi-particle features far away from these quantum critical
points, around optimum doping ($p=p_{m}$, see Fig.\ref{fig1}), can
be understood. Here quantum fluctuations do not dominate. Yet
another facet of the mystery emerges from the evidence for a
finite size effect in the temperature dependence of the in-plane
penetration depth. Although the limiting length scales may depend
on the history of the sample, their dependence on oxygen isotope
substitution reveals and confirms the coupling between
superconducting properties and lattice degrees of freedom.
Although the majority opinion on the mechanism of
superconductivity in the cuprates is that it occurs via a purely
electronic mechanism, and lattice degrees of freedom are supposed
to be irrelevant, we have shown that the oxygen isotope effect on
the in-plane penetration depth uncovers yet another facet, the
hitherto ignored coupling between lattice degrees of freedom and
both, normal state and superconducting state properties, the
existence of inhomogeneities giving rise to a finite size effect.
Finally we note that a variety of other properties also display
pronounced phonon and electron-lattice effects:
superconductivity-induced lattice changes\cite
{pasler,meingast,you,chmaiss,sani}, superconductivity-induced
phonon
renormalization\cite{chung,pintschovius,egami,hadjiev,mook,lanzaraph},
tunnelling phonon structures\cite{shimada,gonnelli}, etc., give
additional evidence of significant electron-lattice coupling.
Furthermore we have seen that the occurrence of power law terms in
the low temperature and low magnetic field dependence of the
in-plane penetration depth do not necessarily single out d-wave
pairing, but stem from fluctuations at work, associated with a
complex scalar order parameter which is compatible with both, s-
and d-wave pairing.

\acknowledgments The author is grateful to M. Cohen, R. Khasanov, H. Keller,
K.A. M\"{u}ller and D. R. Nelson for very useful comments and suggestions on
the subject matter, and to Claudia Ambrosch-Draxl and R. Zeyher for
clarifying correspondence.


\begin{references}

\bibitem{tallon}  J. L. Tallon {\em et al.}, Phys. Rev. B {\bf 51}, 12911
(1995).

\bibitem{presland}  M.R. Presland {\em et al.}, Physica C {\bf 176}, 95
(1991).

\bibitem{tseuro}  T. Schneider, Europhys. Lett. {\bf 60}, 141 (2002).

\bibitem{tsphysB}  T. Schneider, Physica B {\bf 326}, 289 (2003).

\bibitem{chandra}  B. S. Chandrasekhar and D. Einzel, Annalen der Physik
{\bf 2}, 535 (1993).

\bibitem{hirschfeld}  P. J. Hirschfeld and N. Goldenfeld, Phys. Rev. B {\bf 48}, 4219 (1993).

\bibitem{durst}  A. C. Durst and P. A. Lee, Phys. Rev. B {\bf 62}, 1270,
(2000).

\bibitem{hardy}  W. N. Hardy, D. A. Bonn, D. C. Morgan, R. Liang, and K.
Zhang, Phys. Rev. Lett. {\bf 70}, 3999 (1993).

\bibitem{tsjsup}  T. Schneider1 and J. M. Singer, J. of Superconductivity,
{\bf 13}, 789 (2000).

\bibitem{martindale}  J. A. Martindale, S. E. Barrett, K. E. OHara, C. P.
Slichter, W. C.Lee, and D. M. Ginsberg, Phys. Rev. B {\bf 47}, 9155 (1993).

\bibitem{lee}  P. A. Lee, Phys. Rev. Lett. {\bf 71}, 1887 (1993).

\bibitem{graf}  M. J. Graf, S-K. Yip, J. A. Sauls, and D. Rainer, Phys. Rev.
B {\bf 53}, 15147 (1996).

\bibitem{taillefer}  L. Taillefer, B. Lussier, R. Gagnon, K. Behnia, and H.
Aubin, Phys. Rev. Lett. {\bf 79}, 483 (1997).

\bibitem{moler}  K. A. Moler, D. L. Sisson, J. S. Urbach, M. R. Beasley, A.
Kapitulnik, D. J. Baar, R. Liang, and W. N. Hardy, Phys. Rev. B {\bf 55},
3954\ (1997).

\bibitem{wright}  D. A. Wright, J. P. Emerson, B. F. Woodfield, J. E.
Gordon, R. A. Fisher, and N. E. Phillips, Phys. Rev. Lett. {\bf 82}, 1550
(1999).

\bibitem{ywang}  Y. Wang {\em et al.}, Phys. Rev. B {\bf 63}, 094508 (2001).

\bibitem{tinkham}  M. Tinkham, {\it Introduction to Superconductivity},
McGraw Hill, New York 1975.

\bibitem{zhaobi}  G. Zhao, V. Kirtikar, and D.E. Morris, Phys. Rev. B {\bf 63}, 024503(2001).

\bibitem{hofer}  J. Hofer {\em et al.}, Phys. Rev. Lett. {\bf 84}, 4192
(2000).

\bibitem{khasanovpr}  R. Khasanov {\em et al.},J. Phys. Condensed Matter
{\bf 15}, L17 (2003).

\bibitem{khasanovd}  R. Khasanov {\em et al.}, cond-mat/0305477.

\bibitem{tallonqcp}  J. L. Tallon {\em et al.}, cond-mat/0211071.

\bibitem{uemura}  Y. J. Uemura, Solid State Communications {\bf 126}, 2338 (2003)

\bibitem{dicastro}  D. DiCastro {\em et al.}, cond-mat/0307330.

\bibitem{picket}  W. E. Pickett, Rev. Mod. Phys. {\bf 61}, 433 (1989).

\bibitem{campuzano}  J.C. Campuzano, M.R. Norman, and M. Randeria,
cond-mat/0209476.

\bibitem{lanzara}  A. Lanzara {\em et al.}, Nature {\bf 420}, 511 (2001).

\bibitem{sflee}  S. F. Lee {\em et al.}, Phys. Rev. B {\bf 53}, 11825 (1996).

\bibitem{conderla}  K. Conder {\em et al.}, in {\it Phase Separation in
Cuprate Superconductors}, edited by E. Sigmund and K. A. Muller (Springer,
Berlin 1994) p. 210.

\bibitem{raffa}  F. Raffa {\em et al.}, Phys. Rev. Lett. {\bf 81}, 5912
(1998).

\bibitem{wang}  N. L. Wang {\em et al.}, Phys. Rev. Lett. {\bf 89}, 087003 (2002).

\bibitem{biggs}  T. Biggs {\em et al.}, cond -mat/0203392.

\bibitem{me}  A. B. Migdal, Sov. Phys. JETP {\bf 7}, 996 (1958); G. M.
Eliashberg, Sov. Phys. JETP {\bf 11}, 696 (1960).

\bibitem{maksimov}  E.I. Maksimov, Zh. Eksp.Teor. Fiz. {\bf 37}, 1562 (1969).

\bibitem{carbotte}  J. P. Carbotte, Rev. Mod. Phys. {\bf 62}, 1027 (1990).

\bibitem{randeria}  A. Paramekanti, M. Randeria, and N. Trivedi,
cond-mat/0305611.

\bibitem{deppeler}  A. Deppeler and A. J. Millis, Phys. Rev. B {\bf 65},
100301 (2002).

\bibitem{deppeler2}  A. Deppeler and A. J. Millis, Phys. Rev. B {\bf 65},
224301 (2002).

\bibitem{ffh}  D.S. Fisher, M.P.A. Fisher, and D.A. Huse, Phys. Rev. B {\bf 43}, 130 (1991).

\bibitem{tshkws}  T. Schneider and H. Keller, Int. J. Mod. Phys. B {\bf 8},
487 (1993).

\bibitem{ohl}  N. Overend, M.A. Howson, and I.D. Lawrie, Phys. Rev. Lett.
{\bf 72}, 3238 (1994).

\bibitem{hub}  M. A. Hubbard {\em et al}., Physica C {\bf 259}, 309 (1996).

\bibitem{kamal}  S. Kamal {\em et al}., Phys. Rev. Lett. {\bf 73}, 1845
(1994).

\bibitem{kamal2}  S. Kamal {\em et al}., , Phys. Rev. B {\bf 58}, 8933
(1998).

\bibitem{pasler}  V. Pasler {\em et al}., Phys. Rev. Lett. {\bf 81}, 1094
(1998).

\bibitem{tsjh}  T. Schneider, J. Hofer, M. Willemin, J.M. Singer, and H.
Keller, Eur. Phys. J. B {\bf 3}, 413 (1998).

\bibitem{tsjs}  T. Schneider and J. M. Singer, Physica C {\bf 313}, 188
(1999).

\bibitem{jhts}  J. Hofer {\em et al}., Phys. Rev. B {\bf 60}, 1332 (1999, B {\bf 62}, 631 (2000)

\bibitem{tsjshou}  T. Schneider and J. S. Singer, Physica C {\bf 341-348},
87 (2000)

\bibitem{book}  T. Schneider and J. M. Singer, {\it Phase Transition
Approach To High Temperature Superconductivity}, Imperial College Press,
London, 2000.

\bibitem{meingast}  C. Meingast {\em et al}., Phys. Rev. Lett. {\bf 86},
1606 (2001).

\bibitem{osborn}  K. D. Osborn, D. J. Van Harlingen, Vivek Aji, N.
Goldenfeld, S. Oh, and J. N. Eckstein, cond-mat/0204417

\bibitem{loram}  J. W. Loram {\em et al.}, Physica C {\bf 235-240}, 134
(1994).

\bibitem{tstclaa}  T. Schneider, Phys. Rev. B {\bf 67}, 134514 (2003).

\bibitem{crowell}  P.A. Crowell, F.W. van Keuls, J.R. Reppy, Phys. Rev. B
{\bf 55}, 12620 (1997).

\bibitem{tshk}  T. Schneider and H. Keller, Phys. Rev. Lett. {\bf 82}, 4899
(2001).

\bibitem{fsize}  M. E. Fisher and M. N. Barber, Phys. Rev. Lett. {\bf 28}
1516 (1972); M. E. Fisher, Rev. Mod.Phys. {\bf 46}, 597 (1974); V. Privman,
{\it Finite Size Scaling and Numerical Simulation of Statistical Systems},
World Scientific, Singapore 1990.

\bibitem{schultka}  N. Schultka and E. Manousakis, Phys. Rev. Lett. {\bf 75}, 2710 (1995).

\bibitem{jacobs}  T. Jacobs, S. Sridhar, Q. Li, G. D. Gu, N. Koshizuka,
Phys. Rev. Lett. {\bf 75}, 4516 (1995)

\bibitem{leem}  S. F. Lee {\em et al.}, Phys. Rev. Lett. {\bf 71}, 3862
(1993)

\bibitem{bled}  T. Schneider, cond-mat/0306668.

\bibitem{liu}  J. Liu, J. Wan, A. Goldman, Y. Chang and P. Jiang, Phys.
Rev.Lett. {\bf 67}, 2195 (1991).

\bibitem{chang}  A. Chang, Z. Rong, Y. Ivanchenko, F. Lu and E. Wolf, Phys.
Rev. B {\bf 46}, 5692 (1992).

\bibitem{cren}  T. Cren, D. Roditchev, W. Sacks, J. Klein, J.-B. Moussy, C.
Deville-Cavellin, and M. Lagu\"{e}s, Phys. Rev. Lett. {\bf 84}, 147 (2000).

\bibitem{lang}  K. M. Lang, V. Madhavan, J. E. Hoffman, E. W. Hudson, H.
Eisaki , S. Uchida and J. C. Davis, Nature {\bf 415}, 413 (2002).

\bibitem{tsrkhk}  T. Schneider, R. Khasanov, K. Conder and H. Keller,
cond-mat/0302403.

\bibitem{chung}  J.-H. Chung {\em et al.}, Phys. Rev. B {\bf 67}, 014517
(2003).

\bibitem{yip}  S. K. Yip and J. A. Sauls, Phys. Rev. Lett. {\bf {69}}, 2264
(1992).

\bibitem{maeda}  A. Maeda, Y. Iino, T. Hanaguri, N. Motohira, K. Kishio, and
T.Fukase, Phys. Rev. Lett. {\bf 74}, 1202 (1995); A. Maeda, T. Hanaguri, Y.
Iino, S. Masuoka, Y. Kakata, J. Shimoyama, K. Kishio, H. Asaoka, Y.
Matsushita, M. Hasegawa, and H. Takei, J. Phys. Soc. Jpn. {\bf {65}}, 3638 \
(1996).

\bibitem{carrington}  A. Carrington, R. W. Giannetta, J. T. Kim, and J.
Giapintzakis, Phys. Rev. B {\bf {59}}, 14 173 (1999).

\bibitem{bidinosti}  C. P. Bidinosti, W. N. Hardy, D. A. Bonn, and Ruixing
Liang, Phys. Rev. Lett. {\bf {83}}, 3277 (1999).

\bibitem{sonier}  J. E. Sonier, J. H. Brewer, R. F. Kiefl, G. D. Morris, R.
I. Miller, D. A. Bonn, J. Chakhalian, R. H. Heffner, W. N. Hardy, and R.
Liang, Phys. Rev. Lett. {\bf {83}}, 4156 (1999).

\bibitem{you}  H. You, U. Welp and Y. Fang, Phys. Rev. B {\bf {43}}, 3660
(1991).

\bibitem{sharapov}  S. G. Sharapov, V. P. Gusynin and H. Beck, Phys. Rev. B
{\bf {66}}, 012515 (2002).

\bibitem{chen}  S. J. Chen, {\em et al.}., Phys. Rev. B {\bf {58}}, 14573
(1998).

\bibitem{panagop}  C. Panagopoulos, T. Xiang, W. Anukool, J. R. Cooper, Y.
S. Wang, and C. W. Chu, Phys. Rev. B {\bf 67}, 220502 (2003).

\bibitem{mpaf}  M. P. A. Fisher, {\em et al.}, Phys. Rev. Lett. {\bf 64},
587 (1990).

\bibitem{cha}  M. C. Cha, M. P. A. Fisher, S. M. Girvin, M. Wallin, P. A.
Young, Phys. Rev. B {\bf 44},6883 (1991).

\bibitem{herbut}  I. F. Herbut, Phys. Rev. B {\bf 61}, 14723 (2000).

\bibitem{herbutqsn}  F.Herbut, Phys.Rev.Lett.{\bf 85}, 1532 (2000).

\bibitem{chmaiss}  O. Chmaissem, J. D. Jorgensen, S. Short, A. Knizhnik, Y.
Eckstein and H. Shaked, Nature {\bf 397}, 45 (1999).

\bibitem{sani}  N. L. Sani, A. Bianconi and A. Oyanagi, J. Phys. Soc. Jap.
{\bf 70}, 2092 (2001).

\bibitem{pintschovius}  L. Pintschovius and W. Reichardt, in Physical
Properties of High Temperature Superconductors IV, edited by D. Ginsberg,
(World Scientific, Singapore, 1994), p. 295.

\bibitem{egami}  T. Egami and S. J. L. Billinge, in Physical Properties of
High Temperature Superconductors V, edited by D. Ginsberg, (World
Scientific, Singapore, 1996), p. 265.

\bibitem{hadjiev}  V.G. Hadjiev, X.J. Zhou, T. Strohm, M. Cardona, Q.M. Lin,
and C.W. Chu, Phys. Rev. B {\bf 58}, 1043 (1998).

\bibitem{mook}  H. A. Mook and F. Dogan, Nature (London) {\bf 401}, 145
(1999).

\bibitem{lanzaraph}  A. Lanzara, P. V. Bogdanov, X. J. Zhou, S. A. Keller,
D. L. Feng, E. D. Lu, T. Yoshida, H. Eisaki, A. Fujimori, K. Kishio,
J.-Shimoyama, T. Noda, S. Uchida, Z. Hussain, and Z.-X. Shen, Nature
(London) {\bf 412}, 510 (2001).

\bibitem{shimada}  D. Shimada, Y. Shiina, A. Mottate, Y. Ohyagi, and N.
Tsuda, Phys. Rev. B {\bf 51}, R16495 (1995).

\bibitem{gonnelli}  R.S. Gonnelli, G.A. Ummarino and V.A. Stepanov, Physica
C {\bf 275}, 162 (1997).

\end{references}
\end{document}